\documentclass[12pt]{iopart}

\usepackage{iopams}  
\usepackage[
    natbib=true,
    style=numeric,
    sorting=none
]{biblatex}
\usepackage[pdftex]{graphicx}
\graphicspath{{figures/}}
\usepackage{tabularx}
\usepackage[table,xcdraw]{xcolor}

\begin{document}

\title[Energetic and Control Trade-offs in Spring-Wing Systems]{Energetic and Control Trade-offs in Spring-Wing Systems}

\author{J Lynch$^1$, E S Wold$^2$,  J Gau$^3$, S N Sponberg$^2$, N Gravish$^1$}

\address{$^1$ Department of Mechanical \& Aerospace Engineering, University of California, San Diego}
\address{$^2$ Department of Physics, Georgia Institute of Technology}
\address{$^3$ Georgia Institute of Technology}
\ead{ngravish@ucsd.edu}

\begin{abstract}
Flying insects are thought to achieve energy-efficient flapping flight by storing and releasing elastic energy in their muscles, tendons, and thorax. 
However, flight systems consisting elastic elements coupled to nonlinear, unsteady aerodynamic forces also present possible challenges to generating steady and responsive wing motions. 
In previous work, we examined the resonance properties of a dynamically-scaled robophysical system consisting of a rigid wing actuated by a motor in series with a spring, which we call a spring-wing system \cite{Lynch2021-ri}. 
In this paper, we seek to better understand the effects of perturbations on resonant systems via a non-dimensional parameter, the Weis-Fogh number.
We drive a spring-wing system at a fixed resonant frequency and study the response to an internal control perturbation and an external aerodynamic perturbation with varying Weis-Fogh number. 
In our first experiments, we provide a step change in the input forcing amplitude and study the wing motion response. 
In our second experiments we provide an external fluid flow directed at the flapping wing and study the perturbed steady-state wing motion. 
We evaluate results across the Weis-Fogh number, which describes the ratio of inertial and aerodynamic forces and the potential energetic benefits of elastic resonance. 
The results suggest that spring-wing systems designed for maximum energetic efficiency also experience trade-offs in agility and stability as the Weis-Fogh number increases. 
Our results demonstrate that energetic efficiency and wing maneuverability are in conflict in resonant spring-wing systems suggesting that mechanical resonance presents tradeoffs in insect flight.
\end{abstract}

%
\vspace{2pc}
%
\submitto{\BB}
%
%
%

\section{Introduction}
Flapping flight is an extremely power-intensive mode of locomotion, requiring both high frequency wingbeats and large forces to produce lift and perform agile maneuvers. 
Flying insects achieve efficient flight through a combination of specialized flight muscles \cite{Josephson2000-qh} and elastic energy storage in the thorax \cite{Weis-Fogh1960-kz, Weis-fogh1973-mm,Dickinson1995-jt}. 
The insect flight system can thus be described as muscle actuation of an elastic structure which oscillates wings to generate aerodynamic forces. 
We call this combination of elastic, inertial, and aerodynamic mechanisms a ``spring-wing'' system \cite{Lynch2021-ri}. 
While significant research focus has been devoted to the aerodynamic force generation of flapping wings, only a few studies have focused on understanding the implications of elastic energy storage and return for flight dynamics and control \cite{Weis-fogh1973-mm,Ellington1999-hp,Dickinson1995-jt,Jafferis2016-ij}.

A consequence of ``spring-wing'' dynamics is that there exist several resonant wingbeat frequencies at which different forms of optimality (maximum amplitude, lift, or efficiency, for example) are achieved \cite{Pons2022-qj}.
Operating at a resonant frequency that maximizes lift can enable significant performance advantage, allowing insects to use smaller muscle forces to achieve particular set of flapping kinematics.
Indeed, roboticists designing insect-scale flapping robots have found that incorporating elasticity and operating near resonance enables higher lift and greater payloads \cite{Baek2009-fe,Ma2015-vs,Jafferis2016-ij,Tu2020-jx,Chen2019-yv}.
However, there are trade-offs inherent in operating at resonance \cite{Jafferis2016-ij, robobeefrequencypaper}. 
For example, if the output wing amplitude is maximized at the resonant frequency, any deviation from that frequency peak will result in a decrease in wing amplitude and require more energy input per wingstroke, thus potentially limiting the flapper's ability to quickly change wing kinematics.
This means that a system with a sharp resonant peak has the benefit of improved efficiency at resonance, but would require larger torques or a longer time to change flapping amplitude or create wingbeat asymmetries to adjust the direction of flapping thrust.
These trade-offs are likely to be different for insects or robots with different physical characteristics, but it's not immediately clear how they scale because of the nonlinearity of flapping aerodynamics \cite{Lynch2021-ri}.

To classify the relative importance of resonance in spring-wing systems we use the Weis-Fogh number ($N$), a dimensionless parameter that describes the ratio between peak inertial and aerodynamic torques.
Assuming roughly sinusoidal wingstroke kinematics, $N$ can be expressed in terms of overall wing inertia $I$, mean aerodynamic drag torque coefficient $\Gamma$, and wingstroke amplitude $\phi_0$:
\begin{equation}
    N =\frac{\max{(\tau_{inertia}})}{\max{(\tau_{aero})}}= \frac{I}{\Gamma \phi_0} 
\end{equation}
Previous work has demonstrated that $N$ governs how much energy can be recovered into the elastic system of insects and robots on each wingstroke, and thus is a measure of resonant efficiency for flapping wing systems.
However, stable and agile flight requires much more than just steady-amplitude wing oscillations, leading to the question at hand: how do spring-wing resonant dynamics impact other aspects of flight such as control and robustness?

We hypothesize that the resonant efficiency of a spring-wing system also influences the insect's flapping dynamics in response to internal control changes and external perturbations. 
We motivate this with several simple thought experiments.
First, an insect with a larger wing inertia would have to put more energy in to driving its wing to full amplitude flapping motion, and the greater momentum of the wing would resist changes to that amplitude (Fig. \ref{fig:transient_intro_concept}b).
Thus, the insect with a larger $N$ would tend to be more sluggish than an insect with lower $N$, and thus lower wing inertia.
On the other hand, consider if the insect is flying in a crosswind and needs to maintain its wingbeat amplitude to maintain stable hovering (Fig. \ref{fig:transient_intro_concept}c).
This is a perturbation that acts aerodynamically on the wing and may cause the wing motion to deviate from steady-state if the aerodynamic perturbation significantly overcomes the momentum of the wing motion.
In the case of aerodynamic perturbations, a higher $N$ where inertia dominates over aerodynamic forces would be less susceptible to wingstroke deviation.
In summary, we hypothesize that the Weis-Fogh ratio is a governing parameter of both wingbeat response timescale (increases with $N$), and susceptibility to aerodynamic perturbations (decreases with $N$). 
These two performance metrics impact maneuverability and stability in competing ways, and thus present a potential dilemma for spring-wing resonant flight.

We set out to study the effects of Weis-Fogh number on 1) the responsiveness of a flapping system to control inputs, i.e. starting from stop or changing amplitude, and 2) the ability of a flapper to maintain flapping kinematics when subjected to an asymmetrical aerodynamic perturbation.
The following section sets out expectations based on an analytically tractable (linear) version of the spring-wing system with a viscous damper in place of aerodynamic drag.
Then, we discuss two experiments on a dynamically-scaled spring-wing robot that measure, respectively, the time it takes for systems with different $N$ to flap up to full amplitude and the ability of those systems to maintain sinusoidal flapping kinematics in the presence of a constant flow perturbation.
Finally, we discuss the implications of these and prior results for the biomechanics of insect flight systems and the design of flapping-wing micro aerial vehicles (FWMAVs).

\section{A motivating example}
\subsection{The Weis-Fogh number governs spring-wing resonance dynamics}
\begin{figure}[t]
    \centering
    \includegraphics[width = \textwidth]{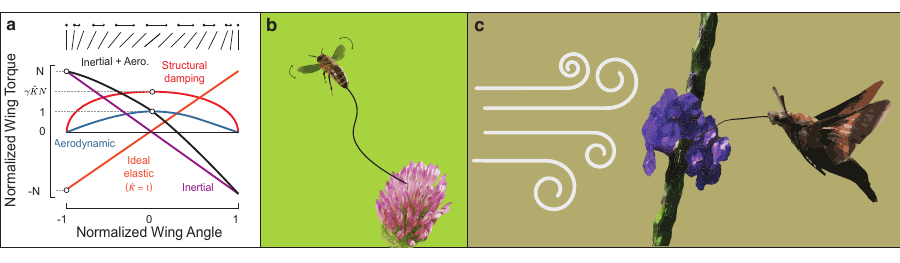}
    \caption[Weis-Fogh number and flight performance]{Weis-Fogh number and flight performance. a) A Weis-Fogh diagram, which illustrates the relationship between inertial, aerodynamic, elastic, and internal damping forces over a wing stroke \cite{Weis-fogh1973-mm, Lynch2021-ri}. The Weis-Fogh number is labeled on the vertical axis. We hypothesize that increasing Weis-Fogh number correlates more sluggish take off from rest (b) and better resiliencec to aerodynamic perturbations (c).}
    \label{fig:transient_intro_concept}
\end{figure}

The Weis-Fogh number is named for Torkel Weis-Fogh, a pioneer in insect flight biomechanics and discoverer of the elastic protein resilin \cite{Weis-Fogh1960-kz}, and it is defined as the ratio between maximum inertial and maximum aerodynamic torque during flapping:
\begin{equation}
    N = \frac{\max(\tau_{inertia})}{\max(\tau_{aero})}
    \label{eqn:N_def}
\end{equation}
Inertial torques are due to the acceleration of the mass of the wing and the surrounding air (added mass or ``virtual'' inertia \cite{Ellington1984-qg}) as the wing flaps, and the aerodynamic torques are due to drag on the wing in the wing stroke plane.
If a spring-wing system has inertia $I$, aerodynamic drag coefficient $\Gamma$, and flaps sinusoidally with peak-to-peak amplitude $\theta_o$, we can express $N$ as $N = I/(\Gamma \theta_o)$ \cite{Lynch2021-ri}.
Weis-Fogh introduced this term as a part of an argument about the necessity of elastic energy storage and return in the flight system of insects \cite{Weis-fogh1973-mm}.
It expresses the relative influence of inertial and aerodynamic effects on the dynamics of a flapping wing; $N < 1$ means aerodynamic forces dominate, whereas $N>1$ means that inertial force is dominant (See Fig. \ref{fig:transient_intro_concept}a).

We found, through dimensional analysis and dynamically-scaled robotic experiments, that it also has a significant relationship to the resonant characteristics of spring-wing systems. 
We can write an equation of motion of a spring-wing system with structural (frequency-independent) damping as defined in \cite{Lynch2021-ri}:
\begin{equation}
    I_t\ddot{\theta} + k\theta + \frac{k \gamma}{\omega}\dot{\theta} + \Gamma |\dot{\theta}|\dot{\theta} = \tau_{in} 
    \label{eqn:spring_wing_dim}
\end{equation}
\noindent
where $I_t$ is the total inertia of the wing plus added mass inertia, $k$ is the spring stiffness, $\gamma$ is the structural damping loss modulus \cite{Beards1995-yp}, $\omega$ is the forcing frequency, and $\tau_{in}$ is the input torque. 
The system constitutes a forced harmonic oscillator with quadratic damping coefficient $\Gamma$, which is typically much larger than the frequency-independent structural damping term \cite{Gau2019-ir}. 
The expression can also be written in non-dimensional form in terms of the dimensionless angular displacement $q$, the non-dimensional stiffness $\hat{K} = \omega_n^2 / \omega^2$ (which is 1 when the system is driven at its natural frequency), the structural damping factor $\gamma$, and the Weis-Fogh number $N$ \cite{Lynch2021-ri}:
\begin{equation}
    \ddot{q} + \hat{K} q + \hat{K} \gamma \dot{q} + N ^{-1}|\dot{q}|\dot{q} = \tilde{\tau}_{in}
    \label{eqn:spring_wing_nondim}
\end{equation}
Previously we found that the dynamic efficiency when flapping at resonance, a measure of the amount of muscle work that goes directly to producing lift/overcoming drag, $\eta = \frac{W_{aero}}{W_{total}}$, scales with $N$ in systems with \textit{any} internal damping losses \cite{Lynch2021-ri} 
\begin{equation}
    \eta =  \frac{1}{1 + \frac{3\pi}{8} \gamma N}    
\end{equation}
Therefore, while it is beneficial to have an $N>1$ for elastic energy exchange and resonance, higher values of $N$ have diminishing returns in terms of peak efficiency.


\subsection{Linear system analysis highlights stability and maneuverability trade-offs in resonant spring-wing flight}
To gain insight into how we should expect the spring-wing to behave in our start up and constant aerodynamic perturbation experiments, we start by studying the behavior of a linear spring-mass-damper.
We choose to use the linear equations because the quadratic aerodynamic damping in the spring-wing equations prohibit closed-form solutions.
However, we will show that features of the linear system are analogous to the nonlinear version and draw conclusions based on that.

\subsubsection{Normalized linear spring-mass-damper}
Consider the forced linear spring-mass-damper equation:
\begin{equation}
    m\ddot{x}+b\ddot{x}+kx= F_0\sin(\omega t)
    \label{eqn:linear_smd}
\end{equation}
We can normalize the differential equation here by dividing by the mass, creating the standard 2nd-order differential equation.
\begin{equation}
    \ddot{x}+2\xi \omega_n\ddot{x}+\omega_n^2x= F_m\sin(\omega t)
    \label{eqn:linear_smd_norm}
\end{equation}
Where $\omega_n$ is the natural frequency of the system and $\xi$ is the damping ratio, which dictates whether the system is overdamped, critically damped, or underdamped.
The forcing term $F_m$ is defined as $F_m = \frac{F_0}{m}$.
Another way to write \ref{eqn:linear_smd_norm} is by defining the quality factor $Q = \frac{1}{2\xi}$:

\begin{equation}
    \ddot{x}+\frac{\omega_n}{Q}\dot{x}+\omega_n^2x= F_m\sin(\omega t)
    \label{eqn:linear_smd_normQ}
\end{equation}
The quality factor here has a number of interpretations, but for our purposes, it represents the ``steepness'' of the resonance peak.
$Q<0.5$ results in an overdamped system with no peak, whereas high $Q$ results in a tall and narrow resonance curve.

\subsubsection{The time to full amplitude varies linearly with $Q$}
\begin{figure}[h]
    \centering
    \includegraphics{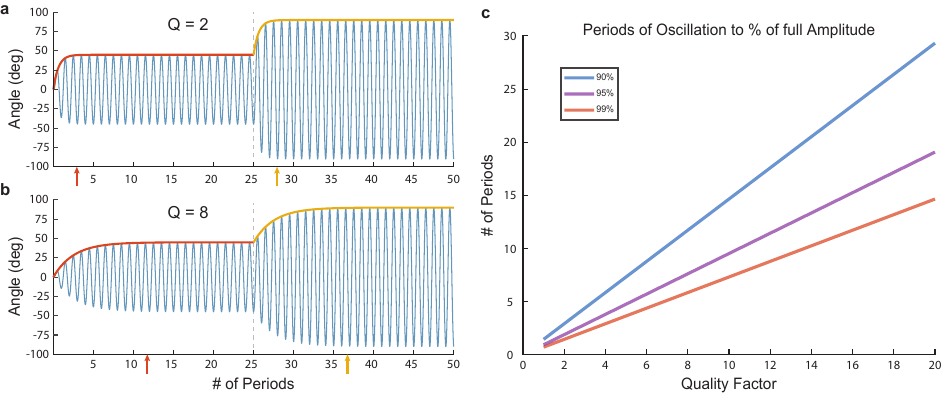}
    \caption[Rise time in linear spring-mass-damper]{Time to full amplitude is proportional to $Q$. a and b show startup to 45 degree amplitude and an amplitude change from 45 to 90 deg for $Q=2$ and $Q=8$ respectively. The rise time is slower with higher Q, as shown in c for several values of $p$. }
    \label{fig:rt_analysis_plots}
\end{figure}

The solution to Eq. \ref{eqn:linear_smd_normQ} for $Q>0.5$ is oscillatory motion that has an initial transient response that depends on initial conditions (which determine oscillation amplitude $X_o$ and phase delay $\delta$), followed by steady-state oscillatory behavior as $t\rightarrow \infty$:
\begin{eqnarray}
    x(t)=X_0 e^{-\lambda t} \cos \left(\omega_{d} t-\delta\right)+R \cos (\omega t - \Delta)
\end{eqnarray}
where 
\numparts
\begin{eqnarray}
    \lambda &= \frac{\omega_n}{2Q} \\
    \omega_d &= \omega_n^2\sqrt{1 - \frac{1}{4Q^2}} \\ 
    \omega_{n} &= \sqrt{\frac{k}{m}}  \\
    R&=\frac{F_m}{\sqrt{\left(\omega_{n}^{2}-\omega^{2}\right)^2+ \frac{\omega_n^2}{Q^2} \omega^{2}}}  \\
    \tan \Delta &= \frac{\omega_n\omega}{Q\left(\omega_{n}^{2}-\omega^{2}\right)}
\end{eqnarray}
\endnumparts
The rate of decay of the transient portion of the ODE solution is
\begin{equation}
    \lambda=\frac{\omega_n}{2Q},
\end{equation}
which is inversely related to $Q$. 
To find the time it takes a forced system to achieve some percentage $p$ of full amplitude, we define a small parameter $\epsilon =1-p$ and solve for $t_p$
\begin{eqnarray}
    \epsilon &= e^{-\lambda_{tr} t_p} \nonumber\\
    \rightarrow t_p &= \frac{-\ln\epsilon}{\lambda_{tr}}\nonumber \\
    &= \frac{-2\ln{\epsilon}}{\omega_n}Q
\end{eqnarray}
or, expressed in terms of the natural period $T_n = 2\pi/\omega_n$
\begin{eqnarray}
    \frac{t_p}{T_n} = \hat{t}_p= \frac{-\ln\epsilon}{\pi} Q 
    \label{eqn:rel_respTime_vsQ}
\end{eqnarray}
Thus we should expect to see that time it takes for transients to decay should be directly proportional to the quality factor $Q$, as shown in Fig. \ref{fig:rt_analysis_plots}.
Notably, $\frac{t_p}{T_n}$ represents a measure of control bandwidth in the spring-wing flapping dynamics. 
An immediate change in oscillatory actuation force ($F_0$) will result a time-lagged ($t_p$) change in the oscillation amplitude, where the time in normal flapping periods $\frac{t_p}{T_n}$ is proportional to $Q$. 
Thus, a spring-wing system with large quality factor will be more ``sluggish'' in response to control input changes, because the response timescale is large.

\subsubsection{The relative influence of aerodynamic perturbations is inversely proportional to $Q$}
Consider a wing flapping in a viscous flow such that the effective velocity at the wing is $\dot{x} - v$.
Ignoring added mass effects that may be present in the aerodynamic system, Equation \ref{eqn:linear_smd_normQ} can be rewritten 
\begin{eqnarray}
    \ddot{x}+\frac{\omega_n}{Q}(\dot{x} - v)+\omega_n^2x= F_m\sin(\omega t) \nonumber\\
    \rightarrow \ddot{x}+\frac{\omega_n}{Q}\dot{x}+\omega_n^2x= F_m\sin(\omega t)+\frac{\omega_n}{Q}v
    \label{eqn:asymmetric_normQ}
\end{eqnarray}
The effect of the perturbation after the transient has decayed is to introduce a torque that biases the spring in the direction of the external flow.
The magnitude of the spring deflection is proportional to the flow velocity and is inversely proportional to $Q$.
Thus the influence of an external flow on a linear flapping system is smaller in a spring-wing system with higher $Q$.

\subsection{Resonance presents competing influences on wing maneuverability and perturbation rejection}

The previous two sections illustrated how the control timescale and susceptibility to aerdynamic perturbations are influenced by the resonant properties of a linear spring-mass-damper. 
The quality factor ($Q$) is an important metric in determining properties of a resonant system, and highlights potential trade-offs in wing maneuverability and stability. 
Higher $Q$ will result in a slower control response from actuation, yet external fluid forces acting on the wing will result in smaller disruption to wing motion. 
Lower $Q$ will result in fast control response from actuation, however external fluid forces will cause disruption to the wingbeat kinematics.
This linear systems analysis provides motivation for examining the role of spring-wing resonance in the timescales of control and susceptibility to aerodynamic perturbations in flapping wing systems. 

\subsection{The Weis-Fogh Number $N$ is the quality factor of a spring-wing system}
One method of comparing the nonlinear spring-wing and linear spring-mass equations is to approximate the linear damping coefficient $b$ with the aerodynamic damping coefficient $\Gamma$ multiplied by the maximum velocity of the wing $\max(\dot{\theta}) = \theta_0 \omega$. 
Defined as such, the damping terms for both spring-mass and spring-wing equations are equivalent at mid-stroke where the wing velocity is highest. 
This is called the secant approximation and has been used in previous analysis of flapping wing systems \cite{Whitney2012-sx}. 
We can define the following relationship for the linear damping coefficient that models the spring-wing
\begin{eqnarray}
    b_{sw} = \Gamma \theta_0 \omega
\end{eqnarray}
Substituting this expression into the equation for the damping ratio yields the following 
\begin{eqnarray}
    \xi &= \frac{b}{2 m \omega_n} \nonumber \\
        &= \frac{\Gamma \theta_0 \omega}{2 m \omega_n}  \nonumber \\
        &= \frac{1}{2N}\frac{\omega}{\omega_n}
\end{eqnarray}
Thus, we see that the Weis-Fogh number has a natural connection to the damping ratio of a linear spring-mass system under the secant approximation. 
If we make the assumption that the system is on resonance ($\omega=\omega_n$) then the relationship is as follows
\begin{eqnarray}
    \xi = \frac{1}{2 N}
\end{eqnarray}
We can push this analogy one step further if we consider how the quality-factor relates to the damping coefficient, and by extension the Weis-Fogh number. 
\begin{eqnarray}
    Q &= \frac{1}{2 \xi} \nonumber \\
      &= \frac{2 N}{2}   \nonumber \\
      &= N
\end{eqnarray}
We have demonstrated that the Weis-Fogh number is equal to the quality factor of a linearized spring-wing system using the secant approximation. 
This corresponds to our measurements in \cite{Lynch2021-ri}.


We test the scaling relationship between Weis-Fogh number $N$ and the dynamic behavior of spring-wings via two experiments in a robophysical model. The first measures response to control inputs by measuring time to peak amplitude from rest, and the second measures the effect of environmental perturbations via measuring the effect of constant cross-flow on symmetry of flapping dynamics.
The results suggest that in addition to its effect on peak dynamic efficiency, $N$ illustrates the scaling of agility and perturbation rejection among insects and other small-scale flapping systems.

\section{Experimental Methods}
In a linear forced spring-mass-damper system, it is possible (as shown above) to derive exact solutions for the transient response of the system during start up or just after a finite perturbation. Those solutions show a clear relationship between the Q-factor, responsiveness, and robustness to environmental effects. We expect that the nonlinear analogue of Q, the Weis-Fogh number $N$, should have a similar relationship to the transient response of the spring-wings system. However, due to aerodynamic drag, the spring-wing equations of motion do not admit an exact solution. To demonstrate the relationship between $N$ and the stability and agility of flapping in spring-wing systems, we perform a series of experiments on a dynamically-scaled robotic spring-wing system. The robotic system is subject to real fluid forces at Reynolds numbers that are scaled to those experienced by insects and insect-scale robots.
\begin{figure}[h]
    \centering
    \includegraphics[width= \columnwidth]{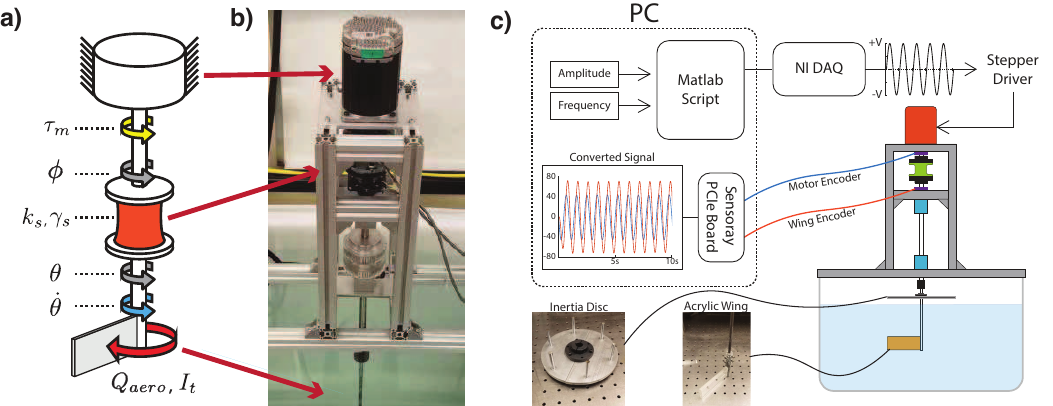}
    \caption[Robotic Spring-Wing System ]{\textbf{The series-elastic spring-wing system.} a) Conceptual diagram indicating the angle input, linear spring with structural damping, and rigid fixed-pitch wing. b) Corresponding photo of the roboflapper indicating the ClearPath (XXXX) servo, silicone torsion spring, and acrylic wing in a large tank of water. c) Diagram of the whole electromechanical system. See \cite{lynch2021-ri} for the full details}
    \label{fig:robo_setup}
\end{figure}

\begin{table}[b]
\centering
\caption{ Inertia and spring stiffness values for the roboflapper}\label{tab:system_parameters}
\begin{tabular}{|cc|lcc}
\cline{1-2} \cline{4-5}
\multicolumn{2}{|l|}{\textbf{Inertia (kg m$^2$)}}                                  & \multicolumn{1}{l|}{} & \multicolumn{2}{l|}{\textbf{Springs (Nm rad$^{-1}$)}}                                                \\ \cline{1-2} \cline{4-5} 
\multicolumn{1}{|c|}{\cellcolor[HTML]{DDEBF7}IA} & \cellcolor[HTML]{DDEBF7}0.00105 & \multicolumn{1}{l|}{} & \multicolumn{1}{c|}{\cellcolor[HTML]{FFF2CC}K1} & \multicolumn{1}{c|}{\cellcolor[HTML]{FFF2CC}0.164} \\ \cline{1-2} \cline{4-5} 
\multicolumn{1}{|c|}{\cellcolor[HTML]{BDD7EE}IB} & \cellcolor[HTML]{BDD7EE}0.00149 & \multicolumn{1}{l|}{} & \multicolumn{1}{c|}{\cellcolor[HTML]{FFE699}K2} & \multicolumn{1}{c|}{\cellcolor[HTML]{FFE699}0.416} \\ \cline{1-2} \cline{4-5} 
\multicolumn{1}{|c|}{\cellcolor[HTML]{9BC2E6}IC} & \cellcolor[HTML]{9BC2E6}0.00233 & \multicolumn{1}{l|}{} & \multicolumn{1}{c|}{\cellcolor[HTML]{FFD966}K3} & \multicolumn{1}{c|}{\cellcolor[HTML]{FFD966}0.632} \\ \cline{1-2} \cline{4-5} 
\multicolumn{1}{|c|}{\cellcolor[HTML]{488FD0}ID} & \cellcolor[HTML]{488FD0}0.00476 &                       & \multicolumn{1}{l}{}                            & \multicolumn{1}{l}{}                               \\ \cline{1-2}
\end{tabular}
\end{table}
\subsection{Dynamically-scaled, series elastic robophysical model}
The robotic spring-wing system used in this paper was described in detail in \cite{Lynch2021-ri} and is shown in Fig. \ref{fig:robo_setup}. It consists of a high-torque servo motor (Teknic ClearPath) connected to a rigid, fixed pitch acrylic wing in a large tank of water. The elasticity comes from a molded silicone torsion spring in series with the wing \ref{fig:robo_setup}. We created three springs from Dragon Skin 30 silicone (SmoothOn) cast in 3D printed molds, varying the geometry so that they each had a different stiffness. We vary the overall inertia of the system by attaching mass to the main shaft of the flapper (above the water) in the form of acrylic and aluminum plates (Fig. \ref{fig:robo_setup}c). We minimize friction by integrating air bearings and thrust ball bearing, and we assume that drag from rotation through the air is much smaller than from the motion of the wing in water. See Table \ref{tab:system_parameters} for a list of the inertia and stiffness values.

\subsection{Controlling $N$, an emergent property of spring-wing flapping systems}
We sought to compare the transient behavior of the flapper when it flaps with different values of $N$. However, due to the dependence on flapping amplitude $\theta_0$, $N$ is an \textit{emergent} quality of a system, and therefore is difficult to prescribe directly. The following section describes the process of determining robotic system configurations for a range of $N = 1-10$ that are used for robustness and agility experiments. In all cases, we refer to a value of $N$ computed using the steady-state flapping amplitude, $\theta_0$, and the mean drag torque coefficient $\Gamma$ at steady state.

\paragraph{Determine constraints} Based on the range of $N$ seen in insects and flapping robots \cite{Weis-fogh1973-mm,Lynch2021-ri}, we sought to test 10 integer values of $N$, from $N = 1$ to $N = 10$. 
Since we are interested in resonant flapping performance, we require the forcing frequency be at the resonant frequency, derived using the method from \cite{Lynch2021-ri}. 
Additionally, we seek to minimize the range of Reynolds number ($Re$) across tests. 
The roboflapper is designed to operate within a range of $Re$ that is similar to insects and small birds ($Re \in [10^2 - 10^4])$, as significant deviations out of that range introduce aerodynamic phenomena that may not be relevant to flapping flight at that scale.

Beyond those considerations, we are limited by constraints on the robotic system. 
Mechanically, we must use one of three silicone springs, one of four discrete inertial configurations, and the same wing with $\Gamma = 1.07\times10^{-3} Nm$. 
On the control side, we found that our system works best when the flapping amplitude is between $\approx$30 and 120$^o$ peak-to-peak and the flapping frequency is between 0.5 and 3 Hz. 

\paragraph{Choosing configurations for values of $N$}
With three springs and four inertia configurations, we have a total of 12 combinations of springs and inertia plates that are possible. 
We have continuous control of the amplitude and frequency within functional bounds.
The process of choosing a configuration for each value of $N$ is as follows:
\begin{enumerate}
    \item Given a particular value of $N$, compute $\theta_o = \frac{I}{\Gamma N}$ for each of the four inertias. Exclude any configuration where $\theta_o > 60^o$
    \item Compute the resonant frequency $f_r$ based on the remaining inertias and the three available springs. Exclude any configurations where $f_r$ is greater than 3 Hz or less than 1 Hz.
    \item Compute the Reynolds number, $Re = \frac{\bar{U}_{tip} \bar{c}}{\nu}$, of flapping based on that amplitude and frequency as well as wing length and chord, 10 cm and 3.6 cm, respectively. Exclude configurations with $Re > \approx 15000$, which is near upper limit of Reynolds number for insects and hummingbirds \cite{Chin2016-ls}.
    \item Select a configuration for each value of $N$ from the non-excluded configurations 
\end{enumerate}
The final selections are given in Table \ref{tab:configs}.

\begin{table*}[]
\centering
\begin{tabular}{|c|l|c|c|c|c|}
\cline{1-1} \cline{3-6}
\multicolumn{1}{|l|}{\textbf{$N$}} &  & \multicolumn{1}{l|}{\textbf{Spring}} & \multicolumn{1}{l|}{\textbf{Inertia}} & \multicolumn{1}{l|}{\textbf{Amplitude (deg)}} & \multicolumn{1}{l|}{\textbf{Frequency (Hz)}} \\ \cline{1-1} \cline{3-6} 
\cellcolor[HTML]{FCE4D6}1        &  & \cellcolor[HTML]{FFF2CC}K1           & \cellcolor[HTML]{DDEBF7}IA            & \cellcolor[HTML]{63BE7B}56.2            & \cellcolor[HTML]{F3E7FF}1.33            \\ \cline{1-1} \cline{3-6} 
\cellcolor[HTML]{FBD9C4}2        &  & \cellcolor[HTML]{FFF2CC}K1           & \cellcolor[HTML]{BDD7EE}IB            & \cellcolor[HTML]{A6D9B5}40.0            & \cellcolor[HTML]{F0E2FE}1.40            \\ \cline{1-1} \cline{3-6} 
\cellcolor[HTML]{F9CEB2}3        &  & \cellcolor[HTML]{FFE699}K2           & \cellcolor[HTML]{DDEBF7}IA            & \cellcolor[HTML]{FCFCFF}18.7            & \cellcolor[HTML]{AA72D4}2.89            \\ \cline{1-1} \cline{3-6} 
\cellcolor[HTML]{F8C2A0}4        &  & \cellcolor[HTML]{FFE699}K2           & \cellcolor[HTML]{BDD7EE}IB            & \cellcolor[HTML]{F7FAFB}20.0            & \cellcolor[HTML]{BC8FDF}2.51            \\ \cline{1-1} \cline{3-6} 
\cellcolor[HTML]{F6B78E}5        &  & \cellcolor[HTML]{FFE699}K2           & \cellcolor[HTML]{9BC2E6}IC            & \cellcolor[HTML]{E3F2E9}25.0            & \cellcolor[HTML]{D2B1EC}2.05            \\ \cline{1-1} \cline{3-6} 
\cellcolor[HTML]{F5AC7B}6        &  & \cellcolor[HTML]{FFD966}K3           & \cellcolor[HTML]{488FD0}ID            & \cellcolor[HTML]{9CD5AC}42.5            & \cellcolor[HTML]{EEDFFC}1.45            \\ \cline{1-1} \cline{3-6} 
\cellcolor[HTML]{F3A069}7        &  & \cellcolor[HTML]{FFD966}K3           & \cellcolor[HTML]{488FD0}ID            & \cellcolor[HTML]{B3DFC0}36.8            & \cellcolor[HTML]{EEDFFC}1.45            \\ \cline{1-1} \cline{3-6} 
\cellcolor[HTML]{F29557}8        &  & \cellcolor[HTML]{FFD966}K3           & \cellcolor[HTML]{488FD0}ID            & \cellcolor[HTML]{C7E7D1}31.8            & \cellcolor[HTML]{DEC4F3}1.80            \\ \cline{1-1} \cline{3-6} 
\cellcolor[HTML]{F08A45}9        &  & \cellcolor[HTML]{FFD966}K3           & \cellcolor[HTML]{488FD0}ID            & \cellcolor[HTML]{D5EDDE}28.3            & \cellcolor[HTML]{DDC3F2}1.81            \\ \cline{1-1} \cline{3-6} 
\cellcolor[HTML]{EE7E32}10       &  & \cellcolor[HTML]{FFD966}K3           & \cellcolor[HTML]{488FD0}ID            & \cellcolor[HTML]{E1F1E8}25.5            & \cellcolor[HTML]{DDC3F2}1.82            \\ \cline{1-1} \cline{3-6} 
\end{tabular}
\caption[Configurations for each value of $N$]{\label{tab:configs} Configurations for each value of Weis-Fogh number. Amplitude is given as half of the peak-to-peak stroke}
\end{table*}

\subsection{Experiment 1: Starting from Rest and Changing Amplitude}
We sought to measure the effect of a system's $N$ on the time it takes for the system to respond to a change in input. 
A straightforward way of doing so is to measure the time it takes for flapping oscillations to reach a steady-state amplitude after startup.
Furthermore, we measured the time it takes to reach a new amplitude after a change in the input.

For each test, the spring stiffness and inertia were set based on the configurations above. 
The system was driven by a sine wave position signal to the servo through Simulink Desktop Real-Time (Mathworks) and a PCIe 6343 interface (National Instruments).
The frequency was set based on the configuration table, but the wing amplitude is not set explicitly because of the series-elasticity of the roboflapper. 
We found previously that modeling does not fully predict the kinematic gain between angular motor amplitude and wing amplitude \cite{Lynch2021-ri}.
Therefore, for each configuration, we found the proper input amplitude to achieve the desired wingbeat amplitude iteratively using a separate Simulink Desktop Real-Time program, prior to the tests, and recorded the input amplitudes. 
When we performed each experiment, we used the input amplitudes to drive the system in open-loop, which was a fairly reliable way to dictate $N$ (See Table \ref{tab:actual_N}).

Each test was performed by starting the sinusoidal position signal and running for 15 flapping periods, long enough for the amplitude to stabilize (Fig. \ref{fig:riseTime}a). 
After 15 periods, the sinusoidal amplitude of the motor position command signal was increased by 50\%, and the experiment continued for a further 15 periods before ending the experiment.
This process was repeated five times for each value of $N$ and sampled at a rate of 1000 samples per period. 
Note that we refer to the number of periods and that each experiment was run at a different frequency (Table \ref{tab:configs}), so the total runtime varied.
The final amplitude $\theta_0$ was determined by fitting a sine curve to the last 5 periods of the each portion of the test using a bounded nonlinear least squares method in MATLAB (Mathworks). 
Then an exponential curve ($f(t) = \theta_0(1-e^{-\lambda t})$ was fit to the peaks of absolute value of the wing angle in the start and step portions of the data, and the time to 95\% of $\theta_0$ was computed using $\hat{t}_{95} = -\ln{(0.05)}\lambda^{-1}f_r$ (Fig. \ref{fig:riseTime}b\&c).
To create the plots, $N$ was recalculated based on actual experimental amplitude measurements, and mean and standard deviation of both $N$ and $\hat{t}_{95}$ were calculated.

\subsection{Experiment 2: Effect of constant cross-flow}
For the second experiment, we wanted to see how $N$ relates to the flapping wing's ability to reject environmental disturbances. 
We did this by subjecting the flapping wing to a constant crossflow and measuring its deviation from a symmetrical sine wave. 
The flow was provided by a submerged aquarium pump (Simple Deluxe LGPUMP400G 400 GPH) fitted with a 1/2" diameter rubber tube. 
The outlet of the tube was positioned such that it was aligned with the acrylic wing in the tank and created the maximal passive deflection (see Figure \ref{fig:constantFlow}a) against the spring, but did not interfere with flapping, i.e. there was no difference between flapping trajectory whether the tube was in place or not. 
We measured the torque on the wing when the pump was on and the flow was perpendicular with the wing.
We found that the torque was approximately 0.01 Nm, only enough to deflect the softest spring about 3.5 degrees.
The maximum peak aerodynamic torque across the experiments is for $N = 1$ and is $\tau_{amax}=\Gamma(\theta_0)^2(2\pi f)^2 = 0.072$~mNm.
Thus the magnitude of the perturbation is significantly lower than the maximum drag induced by flapping motion, but is still enough to induce asymmetry in flapping.

We ran the flapper with a constant sinusoidal input that produced a wing amplitude consistent with the proper configuration at each value of $N$. 
We recorded the wing trajectory with the pump off to set a baseline at each value of $N$, then turned the pump on. 
We analyzed the impact of aerodynamic perturbation on the flapping kinematics by fitting a sine function to the wing trajectory at steady state using MATLAB functions (Mathworks) and recording the fit error (RMSE).
The fit error was normalized to the flapping amplitude at that configuration so that it represents the fraction of flapping amplitude and is unitless.
Additionally, we noted a change in steady flapping amplitude with the pump on, and plotted the relative change in amplitude from the no flow case to the constant flow case.

\section{Results}

\subsection{Start time and step time increase linearly with $N$}
\begin{figure*}
    \centering
    \includegraphics[width=1\linewidth]{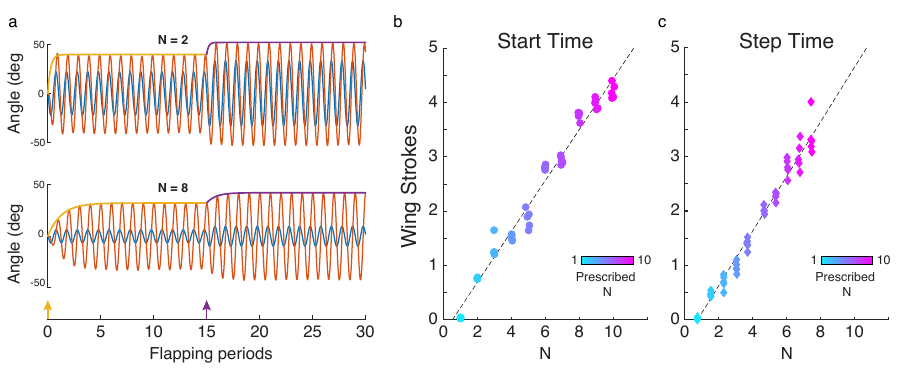}
    \caption[Responsiveness Experiments]{Responsiveness Experiments. a) We drive the series-elastic system via servo (blue) and measure the emergent flapping kinematics (orange). We fit exponential curves to the flapping peaks during start up (yellow) and after an input step (purple) 15 cycles after start. The exponential rates decrease with larger $N$. The measured time (in wing strokes) to full amplitude is linearly related to $N$ (b \& c). However, the effective value for $N$ is less than prescribed after the step due to an increase in flapping amplitude (c).}
    \label{fig:riseTime}
\end{figure*}

We measured the time interval from initiation of flapping to reaching 95\% of the steady-state flapping wing amplitude from an exponential fit.
To normalize this time period across different resonance frequencies we multiplied the start-up time by the frequency of flapping, resulting in a measurement of the the number of wing strokes to reach steady-state. 
The results across $N$ are shown in Fig. \ref{fig:riseTime}b. 
We see that there is a clear relationship between increasing $N$ and increasing time to full amplitude. 
Configurations with $N = 1 \mathrm{~or~} 2$ are at full amplitude within a single wingstroke, whereas $N = 8-10$ systems take four or more wingstrokes. 

The relationship is linear ($t_{start} = 0.486N -0.243$), but there is a small amount of spread both vertically and horizontally (Appendix, Figure \ref{fig:rt_errorbars}) for each value of $N$. 
The vertical spread is to be expected due to fitting in the presence of noise, but we also found that the rise time was sensitive to whether or not the system was at exactly the resonant frequency, maximizing the kinematic gain ($G_k = \theta_o / \theta_{input})$.
This was a more significant issue at higher $N$ because of the steeper resonant curve, and the added mass of the system would have made it susceptible to small asymmetries and potentially larger friction, though that was mediated by thrust ball bearings.
The horizontal spread indicates that we were not always \textit{exactly} at the desired value of $N$. 
The means and standard deviations of $N$ from 1-10 are shown in Table \ref{tab:actual_N}. 

The response time after a step increase in input also has a clear linear relationship with $N$ (Fig. \ref{fig:riseTime}c).
The linear fit is slightly different from the startup data ($t_{start} = 0.503N -0.376$), but they largely fall upon the same line. 
The major difference between the two is that because of the change in amplitude after the step, the effective value of $N$ is lower than it was before the step.
The effect is a compression of the datapoints along the diagonal, since the response time decreases along with $N$, as shown in Figure \ref{fig:discussion_arrows}.

\begin{figure}
    \centering
    \includegraphics[width=0.9\linewidth]{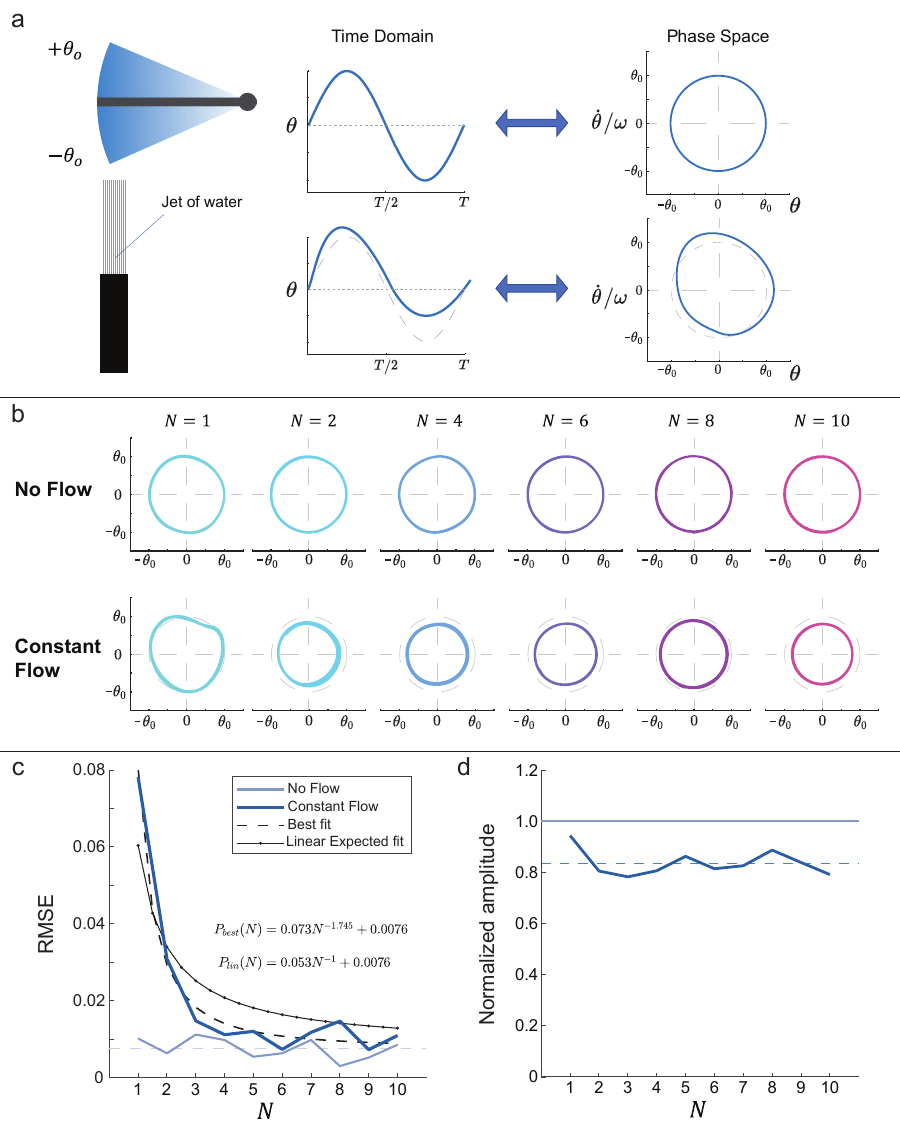}
    \caption[Constant Flow Experiments]{Description of the Constant flow experiments. a) Schematic of the orientation of the water jet relative to the wing, and conceptual representation of the effects of flow on time and phase domain plots. b) Variation in limit cycle plots across $N$. Plots show 2 periods of steady oscillation with and without flow, at different values of $N$. c) Plots of fit error for flow and no-flow cases across $N$, as well as fit lines for expected $N^{-1}$ function and a best fit curve. d) Illustration of relative amplitude decrease across $N$, and mean at 84\% of full amplitude.}
    \label{fig:constantFlow}
\end{figure}

\subsection{Resistance to perturbations increases with increasing $N$}
A flapping wing at steady-state amplitude was subjected to a transverse flow and we measured the change in wingstroke kinematics after flow onset.
We observed that larger $N$ spring-wing systems sustained sinusoidal flapping wing kinematics in the presence of flow in the presence of perturbations, whereas lower $N$ systems exhibited a distortion in the sinusoidal wing motion \ref{fig:constantFlow}.
This asymmetrical warping of the wing trajectory at low $N$ by an external flow was observed as non-sinusoidal wing kinematics (\ref{fig:constantFlow}a) and a non-circular phase portrait (Fig. \ref{fig:constantFlow}a, b).
When there is no flow across the wing, wingstrokes are very close to sinusoidal (circular in phase space) regardless of $N$.
When the flow is turned on, however, the kinematics are affected significantly and deviate from sinusoidal (Fig. \ref{fig:constantFlow}b).
Note that the trajectories are ``lumpier'' when $N$ is small, but also that there's a decrease in flapping amplitude overall at higher $N$.

We fit a sine wave to each trajectory and calculated the root mean squared error (RMSE), which quantifies how well a sine wave fits to the data.
This error will never reach zero, but for the no-flow case, it is small - just 0.76\% of the flapping amplitude.
We find that the sinusoidal fit error at low $N$ is approximately ten times larger than for $N \approx 10$ with a maximum error near 8\% of the flapping amplitude.

Based on Eq. \ref{eqn:asymmetric_normQ} and our association of $Q$ and $N$, we expect that the influence of asymmetric flow should be inversely proportional to $N$.
We fit an inverse curve $AN^{-1}+C$ to the error data, fixing the offset $C = 0.0076$ to be equal to the measured baseline no-flow error.
The optimal curve ($P_{lin}(N)= 0.053 N^{-1}+0.0076$) based on the linear analysis does a fairly poor job of fitting the data ($R^2=0.80$).
Thus we relaxed the constraint on the power of $N$ and fit the curve $AN^{-B}+C$, which produced a curve ($P_{lin}(N)= 0.073 N^{-1.745}+0.0076$) that fit the data much more closely ($R^2=0.97$). 
Noting the reduction in flapping amplitude in the flow cases, we also measured the final amplitude for each $N$ configuration and compared to the amplitude at that value of $N$.
We found that the amplitude was consistently reduced by about 16.4\%, regardless of $N$.
This is unlike the expectation from the linear (viscous) case, where we would not expect to see a decrease in amplitude, just a shift in the center of oscillation.

\section{Discussion}

\subsection{Control authority and Weis-Fogh number}
We have shown that the time it takes for a spring-wing system to respond to a control input change is linearly related to $N$.
Thus a flyer with a greater Weis-Fogh number - determined by their wing mass, wing shape, wing-stroke kinematics, and wing pitch kinematics - will have reduced control authority when it comes to starting, modulating, or stopping wing kinematics.
In order to perform high-speed agile maneuvers, insects need to be able to quickly modulate lift and drag forces, which they can do by modulating amplitude, frequency \cite{Gau2021-fx}, wing rotation \cite{Dickinson1999-ok,Bayiz2021-us}.
However, because of the increased inertial component of a high Weis-Fogh number, control inputs are likely to be damped out to some degree, especially if those inputs come from modulations of the flight muscle.
Therefore, other methods, such as the modulation of wing angle of attack or joint characteristics via steering muscles \cite{Deora2017-fv} may be more effective at high $N$ since they can modulate both amplitude \textit{and} Weis-Fogh number by changing the aerodynamic characteristics of the wing. 

\begin{figure}[t]
    \centering
    \includegraphics{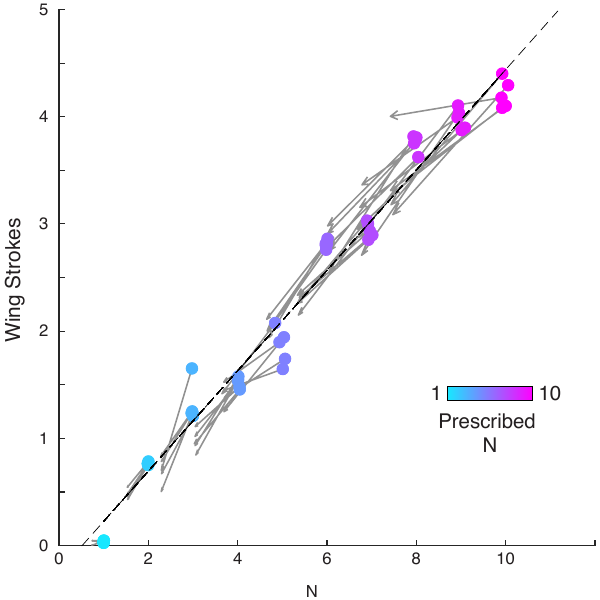}
    \caption[Changes in $N$ and $t_{95}$]{Amplitude increases due to a step increase in input amplitude lead to commensurate decreases in $N$ and transient time. Data points are shown in color, and gray arrows indicate the movement due to increased control input. The arrows are all roughly aligned with the trendline.}
    \label{fig:discussion_arrows}
\end{figure}

Our input step experiments demonstrate that, since $N$ is a function of flapping amplitude, changes to amplitude also change the control authority.
Figure \ref{fig:discussion_arrows} shows the degree to which $N$ and transient time shift due to the increase in amplitude.
The arrows start at the points corresponding to the start up time and point to the location in the plane where the step time is located.
The arrows follow the linear trendline, and the length of the arrow is greater for larger starting values of $N$.

In engineering, control authority is critical for ensuring that a system can meet performance objectives like stabilization in the presence of disturbances and trajectory following.
The loss of control authority - stalling in an airplane, for example - can lead to catastrophic failures if control is not regained in time.
In insects, the ability to quickly maneuver through an array of obstacles or out of the grasp of a predator is similarly important.
Since $N$ is relatively easy to measure for a particular species of insect, requiring just estimates of wing mass, wing shape, and wing kinematic data, it may serve as a useful metric for an insect's relative ability to perform agile maneuvers.

\subsection{Higher $N$ provides greater stability in unpredictable natural environments}
An insect that needs to be more agile may benefit from a lower $N$, but there is a trade off.
At lower $N$, the inertia of the wing during flapping is on the same order as the aerodynamic forces, so environmental flow has a larger effect on flapping kinematics.
This can pose issues for an insect, since steady wingbeats are necessary to produce consistent lift.
Our flow experiments show that an insect or FWMAV with body elasticity is less susceptible to disruptions from the environment when it has a high Weis-Fogh number.
This means that a flapper that needs to fly in a windy environment may benefit from lower amplitude flapping, more massive wings, and/or wing shapes that minimize drag.

\subsection{Weis-Fogh number as the quality factor of spring-wing systems}
The series of analyses we performed looking at the transient behavior of a linear spring-mass-damper as an analogue to the spring-wing system illustrate that the quality factor $Q$ is linearly related to the start up time of the system and inversely related to the relative effect of external perturbations.
Our experimental results with the nonlinear spring-wing system show similar trends, but with some crucial differences

\subsubsection{Changing amplitude changes the transient time constant}
As shown in Fig.\ref{fig:discussion_arrows}, the response time of the spring-wing to a control input depends not just on the magnitude of the input, but also on the amplitude of flapping.
This is an inherently nonlinear phenomenon due to aerodynamic damping, and is not the case for the linear system.
However, since the shift induced by the amplitude change (gray arrows, Fig. \ref{fig:discussion_arrows}) follows the trendline fairly closely, it does seem that the relationship between $N$ and response time is maintained despite the transient changes in $N$.

The actual relationship we expect between response time (defined at 95\% of the full amplitude) and $Q$ based on equation \ref{eqn:rel_respTime_vsQ} is
\begin{equation}
    \hat{t}_{95} = \frac{-\ln{0.05}}{\pi}Q = 0.9536 Q \approx Q
\end{equation}
If we inspect the trendline we fit to the start up time, we find the relationship
\begin{equation}
    \hat{t}_{95} = 0.486N \approx \frac{N}{2}
\end{equation}
Thus it appears that in this nonlinear case, $N$ has the same effect on the transient response rate as $2Q$.

\subsubsection{Nonlinear aerodynamics results in more stability at higher $N$}
In the beginning of this chapter, we argued that an external flow should affect a linear spring-mass-damper less as $Q$ increases, i.e.
\begin{eqnarray}
    F_{flow} = \frac{\omega_n}{Q}v
\end{eqnarray}
Thus we expected an inverse relationship between $N$ and flapping non-sinusoidality.
Additionally, we expected that a flow should cause a consistent off-center stretch in the spring, i.e. a steady-state offset in the positive $x$ direction (Eq. \ref{eqn:asymmetric_normQ}), but maintain the same flapping amplitude.

In fact, we found that an inverse ($N^{-1}$) relationship did not fit the data well.
Instead, a function with $N^{-1.745}$ fit better, suggesting that the quadratic relationship between the system and the flow asymmetry, $\Gamma|\dot{x}-v|(\dot{x}-v)$, introduces dynamics that result in greater passive stabilization of the sinusoidal wing kinematics.
Additionally, we see that the flapping amplitude \textit{is} affected by the asymmetry, causing a decrease in overall amplitude.
This would be detrimental to a high-$N$ flapping flyer's ability to produce lift, but as long as it is not using maximum muscle strength during normal flapping, it should be able to increase the force it uses to drive the wings to achieve the necessary amplitude.
This is in contrast with a low-$N$ flyer, who would need to control amplitude variations within a single wingstroke to maintain smooth flapping, regardless of the strength of the muscle

\subsection{Weis-Fogh number as a performance metric for flapping fliers - living or engineered}
\begin{figure}
    \centering
    \includegraphics[width=0.8\linewidth]{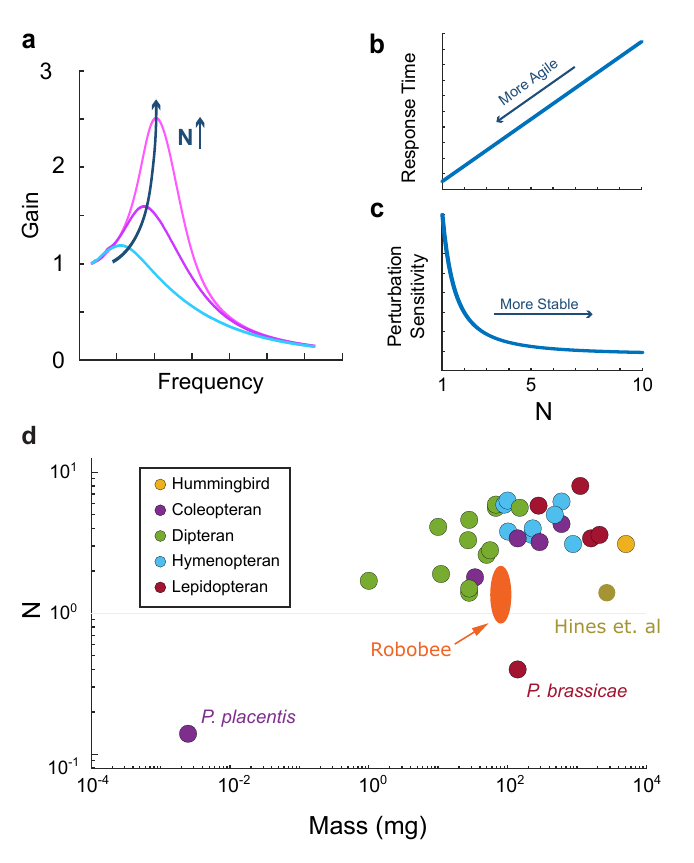}
    \caption[Trade-offs]{Flapping system performance trade-offs. a) Higher $N$ means greater flapping amplitude for a given actuator. b) Lower $N$ leads to faster response times, but c) more vulnerability to aerodynamic perturbations. d) This may point to an explanation for the number of insects and flapping micro-aerial vehicles \cite{Ma2013-ry,Hines2013-yr} across orders of magnitude of size that remain within the range of $N=1-8$. Low-$N$ exceptions like \textit{P. placentis}\cite{Farisenkov2022-sy} and \textit{P. brassicae}\cite{Weis-fogh1973-mm} may point to unique adaptations for efficient flight.}
    \label{fig:discussion_tradeoffs}
\end{figure}

In this and previous work \cite{Lynch2021-ri}, we have shown that the Weis-Fogh number is a metric that encompasses important performance characteristics for flapping flight: dynamic efficiency, responsiveness/agility, and stability.
When we observe the distribution of Weis-Fogh Number across a wide range of insects, we notice that, large or small, they seem to exist in the range of $N=1-8$.
There are some exceptions, of course, but they are characterized by the extremely small flying insects \cite{Farisenkov2022-sy} who fly at very low Reynolds numbers, and butterflies, whose especially large wings and stuttering wing stroke dynamics distinguish them from the more controlled hovering of flies, bees, and hawkmoths.
Small insects like those studied in \cite{Farisenkov2022-sy} likely have values of $N<<1$ (\textit{Paratuposa placentis}, $N\approx 0.14$, see Appendix) due to the presence of bristled wings that significantly decrease wing inertia, and therefore drive $N$ to be smaller; butterflies, like \textit{P. Brassicae, $N\approx0.4$}, also have $N<1$, but via large aerodynamic drag from large wings.
Since the benefits of elastic energy storage drop off when $N<1$, we would expect that such insects need to develop adaptations other than thorax elasticity to maintain flight.
However, thorax elasticity is critical when flight requires wings with significant inertia and high frequency wingbeats.
The fact that other insects who rely on fast wingbeats exist in this constrained range of Weis-Fogh number suggests that the variation in $N$ may reflect a balance of different performance trade-offs (Figure \ref{fig:discussion_tradeoffs})

Similarly, mechanical system parameters can reflect trade-offs between agility and stability.
Fighter aircraft with adjustable wings are one example of a system that can shift from a more stable shape (wings extended) to a faster, more agile, but less stable configuration  (wings folded).
This has been taken to an extreme with fighter jets with forward-swept wings, like the Grumman X-29, which trades off high maneuverability for increased instability.
Indeed, there is even some evidence that wing morphing in birds similarly leverages aerodynamic instability to improve flight performance \cite{Harvey2022-hd}.

Thus it makes sense that the evolutionary development of flapping flight should also balance energetics, agility, and stability.
Perhaps the restriction of flapping animals to a moderate region of Weis-Fogh number ($N=1-8$) is due to tradeoffs that occur between 1) energetic efficiency (increases with $N$), 2) wing stroke responsiveness to control inputs (decreases with $N$), and 3) passive wingstroke stability when subjected to external perturbations (increases with $N$).
Those, combined with the necessity of elastic energy exchange to maintain efficient flight, may constitute a driver of evolutionary change.

\clearpage
\newpage
\appendix
\section{Computing $N$ using measures of flapping power}
When the maximum aerodynamic and inertial torques are not available to compute the Weis-Fogh number, it is also possible to approximate using the aerodynamic and inertial power.
Note, this approximation assumes sinusoidal wingstrokes, which is far from guaranteed; however, this gives a first-order approximation that can be improved through deeper analysis.

Given a sinusoidal wing trajectory $\phi = \phi_o \sin(\omega t)$, the inertial and aerodynamic torques on the wing, according to \ref{eqn:spring_wing_dim}, are
\begin{eqnarray}
    T_i &= I\ddot{\phi} = -I\phi_o\omega^2\sin(\omega t)\\
    T_a &= \Gamma|\dot{\phi}|\dot{\phi} = \Gamma|\phi_o^2\omega^2|\cos(\omega t)|\cos(\omega t)
\end{eqnarray}
The respective inertia and aerodynamic powers are therefore
\begin{eqnarray}
    P_i = T_i\dot{\phi} &=(-I\phi_o\omega^2\sin(\omega t))\phi_o \omega \cos(\omega t) =-0.5I\phi_o^2\omega^3\sin(2\omega t)\\
    P_a = T_a\dot{\phi}&= [\Gamma\phi_o^2\omega^2|\cos(\omega t)|\cos(\omega t)]\phi_o \omega \cos(\omega t) \nonumber\\ 
    &= \Gamma\phi_o^3\omega^3 |\cos(\omega t)|\cos^2(\omega t)
\end{eqnarray}
The maximum magnitudes are $|P_i|_{max} = 0.5I\phi_o^2\omega^3$ and $|P_a|_{max} = \Gamma\phi_o^3\omega^3$. Therefore,
\begin{equation}
    \frac{|P_i|_{max}}{|P_a|_{max}} = \frac{0.5I\phi_o^2\omega^3}{\Gamma\phi_o^3\omega^3} = \frac{I}{2\Gamma\phi_o}=\frac{N}{2}
\end{equation}
and 
\begin{equation}
    N = 2 \times\frac{|P_i|_{max}}{|P_a|_{max}}
\end{equation}

We use this estimate of $N$ to plot the featherwing beetle \textit{Paratuposa placentis} alongside the reported values of $N$ from Weis-Fogh \cite{Weis-fogh1973-mm}. Based on this relationship, we can inspect figure 3e from \cite{Farisenkov2022-sy} and see that there is a maximum (mass specific) aerodynamic power of $\sim$110 W kg$^{-1}$ and inertial power of $\sim$7.8 W kg$^{-1}$. Thus $N \approx 0.14$. We were able to place it on the chart in Fig. \ref{fig:discussion_tradeoffs}e using the fact that the reported body mass is 2.43 $\pm$ 0.19 $\mu$g\cite{Farisenkov2022-sy}. 

It is also possible to compute $N$ from mean values of $P_i$ and $P_a$, as opposed to maxima. In that case, we integrate the expressions for $P_i$ and $P_a$ over the portion of the wingstroke where they are both positive (the first half of the half-stroke):
\begin{eqnarray}
    \bar{P}_i = \int_0^{\frac{\pi}{2\omega}} 0.5I\phi_o^2\omega^3\sin(2\omega t) dt = \frac{1}{2}I\phi_o^2\omega^2 \\
    \bar{P}_a = \int_0^{\frac{\pi}{2\omega}} \Gamma\phi_o^3\omega^3 |\cos(\omega t)|\cos^2(\omega t) = \frac{2}{3} \Gamma \phi_0^3\omega^2
\end{eqnarray}
\noindent
We can then relate $N$ to the ratio of these values:
\begin{equation}
    \frac{\bar{P}_i}{\bar{P}_a} = \frac{\frac{1}{2}I\phi_o^2\omega^2}{\frac{2}{3} \Gamma \phi_0^3\omega^2} = \frac{3}{4}\frac{I}{\Gamma \phi_0} = \frac{3}{4}N
\end{equation}
Thus we can take measurements of mean inertial and aerodynamic power, such as that reported by Ellington \cite{Ellington1984-pp}, and compute $N$ using the relationship $N = \frac{4}{3}\frac{\bar{P}_i}{\bar{P}_a}$. We have included data from \cite{Ellington1984-pp} calculate in this way in Figure \ref{fig:discussion_tradeoffs}.


\section{Additional Tables and Figures}

\begin{figure}[h]
    \centering
    \includegraphics{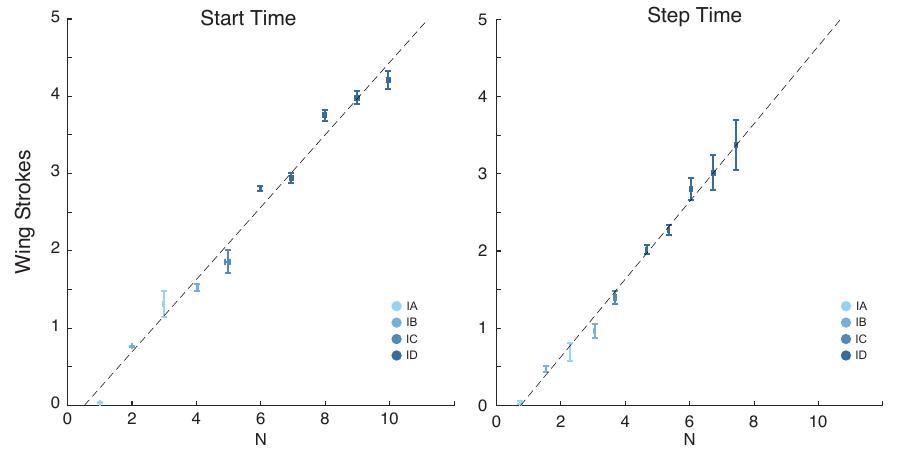}
    \caption{Start and Step Time data plotted with error bars representing standard deviation in measured $N$ (horizontal) and $t_{95}$ (vertical}
    \label{fig:rt_errorbars}
\end{figure}

\begin{table}[h]
    \centering
    \begin{tabular}{|c|c|}
    \hline
    \textbf{Goal N} & \textbf{Actual N} \\ \hline
    1 & $1.0 \pm 0.01$ \\ \hline
    2 & $2.0\pm 0.01$ \\ \hline
    3 & $3.0\pm 0.01$ \\ \hline
    4 & $4.0\pm 0.01$ \\ \hline
    5 & $5.0\pm 0.04$ \\ \hline
    6 & $6.0\pm 0.01$ \\ \hline
    7 & $6.9\pm 0.02$ \\ \hline
    8 & $8.0\pm 0.02$ \\ \hline
    9 & $9.0\pm 0.03$ \\ \hline
    10 & $10.0\pm 0.03$ \\ \hline
    \end{tabular}
    \caption{Mean Experimental $N$ values and standard deviations}
    \label{tab:actual_N}
\end{table}

\begin{table}[h]
    \centering
    \begin{tabular}{|c|c|c|c|c|c|c|c|c|c|c|}
    \hline
    \textbf{Goal N} & 1 & 2 & 3 & 4 & 5 & 6 & 7 & 8 & 9 & 10 \\\hline
    \textbf{Mean N} & $1.0$  & $2.0$ & $3.0$ & $4.0$
                        & $5.0$ & $6.0$ & $6.9$ & $8.0$
                        & $9.0$ & $10.0$  \\ \hline
    \textbf{$\pm$} &$0.01$ &$0.01$ & $0.01$&$0.01$ &$0.04$ 
                     &$0.01$& $0.02$&$0.02$ &$0.03$ & $0.03$ \\ \hline    
    \end{tabular}
    \caption{Mean Experimental $N$ values and standard deviations}
    \label{tab:actual_N_alt}
\end{table}

\newpage
\printbibliography

\end{document}